\documentclass{article}

\usepackage{color, colortbl}
\usepackage{caption}
\usepackage{graphicx}

\definecolor{Gray}{gray}{0.85}
\definecolor{LGray}{gray}{0.96}

\usepackage[top=1in, bottom=1.25in, left=1.25in, right=1.25in]{geometry}
\begin{document}

\title{Benford's Law Applies To Online Social Networks}

\author{Jennifer Golbeck\\
University of Maryland, College Park, Maryland USA}



\maketitle

\begin{abstract}
{Benford's Law states that the frequency of first digits of numbers in naturally occurring systems is not evenly distributed. Numbers beginning with a 1 occur roughly 30\% of the time, and are six times more common than numbers beginning with a 9. We show that Benford's Law applies to social and behavioral features of users in online social networks. We consider social data from five major social networks: Facebook, Twitter, Google Plus, Pinterest, and Live Journal. We show that the distribution of first significant digits of friend and follower counts for users in these systems follow Benford's Law. The same holds for the number of posts users make. We extend this to egocentric networks, showing that friend counts among the people in an individual's social network also follow the expected distribution. We discuss how this can be used to detect suspicious or fraudulent activity online and to validate datasets.  }
\end{abstract}


Benford's Law is an, at first ,unintuitive principle that states that in many naturally occurring systems, the distribution of first digits is not uniform. Numbers beginning with a ``1'' are far more common - more than six times as frequent - than numbers beginning with ``9''. The exact frequency $P$ predicted for a digit $d$ is given by the following formula:
$P(d) = \log_{10}(1+\frac{1}{d})$

Benford's Law is frequently used in forensic accounting, where a distribution of first digits that does not conform to the expected distribution may indicate fraud\cite{durtschi2004effective}. Research has also shown that it applies to genome data\cite{hoyle2002making}, scientific regression coefficients\cite{diekmann2007not}, election data\cite{tam2007breaking,roukema2009benford}, the stock market \cite{hill1998first}, and even to JPEG compression \cite{fu2007generalized}.

We conducted an analysis over five of the most popular social networking websites and showed that Benford's Law applies to the social network structure in all of them. Specifically, the first significant digit (FSD) of users' friend and follower counts on Facebook, Twitter, Google Plus, Pinterest, and LiveJournal all follow Benford's Law. The number of posts that users make also conform to Benford. To our knowledge, this is the first time Benford's Law has been applied to social networks.  We show that exceptions to this rule can uncover configurations within social media systems that lead to unexpected results. 

Furthermore, we show that for any individual, the distribution of friend counts within his or her egocentric network also follows Benford's Law. When the expected distribution is violated, it indicates unusual behavior. A preliminary analysis of over 20,000 Twitter accounts showed that the 100 users whose egocentric networks deviated most strongly from the Benford's Law distribution were all engaged in suspicious activity. 

We discuss how these results lead to the possibility of Benford's Law being used to detect malicious or irregular behavior on social media. We also show that it could be used to validate the sampling in social media datasets. 

\section{Benford's Law: Background and Related Work}

The astronomer Simon Newcomb first formulated what came to be known as Benford's Law in the 1880s. He had noticed that books with logarithm tables showed a lot more wear toward the front, where the numbers beginning with 1 were, than in the back toward the 9s. Concluding that numbers beginning with 1 must be more common, he calculated the probability formula mentioned above.

Physicist Frank Benford noticed the same phenomenon as Newcolm. He validated the observation by collecting naturally occurring numbers from many sources - the surface area of rivers, atomic weights, and numbers appearing in Reader's Digest \cite{benford1938law}. All the values followed the pattern, and while they were not a perfect match \cite{diaconis1979rounding}, the principle of low first digits appearing exponentially more often than high ones was established.

The formula for the law, $P(d) = \log_{10}(1+\frac{1}{d})$, provides a theoretical distribution of expected first digits, shown in Table 1. 
\begin{table*}
\begin{centering}
\begin{tabular}{lrrrrrrrrr} 
\rowcolor{LGray}\textbf{FSD} & 1 & 2 & 3 & 4 & 5 & 6 & 7 & 8 & 9\\
\textbf{Frequency} & 0.301 & 0.176 & 0.125 & 0.097 & 0.07918 & 0.067 & 0.05799 & 0.051 & 0.046\\
\end{tabular}
\caption{Frequency of First Significant Digits (FSD) expected by Benford's Law.}
\end{centering}
\end{table*}
 
On the surface, Benford's Law is quite counter-intuitive. Why would numbers beginning with 1 be any more common than those beginning with 9?  However,  the law holds across many variations in measurement \cite{stoessiger2013benford}. Temperatures that follow Benford's Law do so regardless of whether they are measured in Farenheit, Celsius, or Kelvin. Distances follow whether measured in miles, kilometers, or smoots. 

Most persuasively, Hill provided a proof for the Benford's Law in 1995 \cite{hill1995statistical} . He suggests that skeptics jot down all the numbers that appear on the front pages of several news papers or randomly select data from the Farmer's Almanac as a simple experiment and personal demonstration that the law holds\cite{hill1998first}.
 
 Benford's Law has been shown to describe many naturally occurring sets of numbers. In addition to those already mentioned, there is some research specifically relevant to this paper. It is known that Benford will often apply to systems that follow a power law distribution \cite{pietronero2001explaining}; power laws are commonly found in social network structure\cite{barabasi1999emergence} and  social media\cite{asur2011trends}. While no work has investigated how well Benford's Law describes  social networks (online or offline) or social media, research has shown it describes human behavior online through price distributions in eBay auctions \cite{giles2007benford}.

\section{Data, Data Sets, and Collection}

We analyzed data from five major social networking websites: Facebook, Twitter, Google Plus, Pinterest, and Live Journal.

We collected the number of friends in each network and followers when appropriate. On Google Plus, Twitter, and LiveJournal we also had access to egocentric network data. For each user, we obtained a list of friends and the count of outgoing edges for each of those friends. With this data, we could analyze the distribution of FSDs for an individual person's social network

On Twitter and Pinterest, we also had access to the number of posts each person made. This provides another interesting insight into the general patterns of behavior on social media and whether Benford's Law applies.

We collected some of these datasets ourselves and used other datasets that had been created by others. The following sections detail our process with each network.

\subsection{Facebook}

We accessed user profiles using the Facebook Graph API with requests for friend counts of a numeric Facebook user ID. Once we accessed a user's data, we incremented the user ID by 10,000 and make the next request. If there was no data available for a given user, we incremented the user ID by 1 and tried the next person until we found a match. We collected friend counts for 18,298 users. 

\subsection{Twitter}

 We collected the number of followers and friends (people the user is following) the user had, and the number of people each of those friends were following. This  allowed us to analyze the distribution of FSDs within  egocentric social networks. In addition to this  network data, we collected the number of status updates for each user. 
 Although there are  existing Twitter social network datasets online, we collected our own data in this project in order to work with non-anonymised users so we could later analyze their account activity. 
 
 We accessed data via the Twitter API, using users' numeric Twitter user ID. Our process was to access a user's data and then increase the user ID by 50,000. This gave us a fairly uniform distribution of users, and we were able to collect the data in a reasonable amount of time (over a few weeks) given Twitter's API limits.
 
If a user ID was not linked to an account or if the account was protected, we skipped that user, incremented the ID by 1, and tried the next person. 
 Because we considered counts for friends, followers, and status updates, we only included users in our sample who had at least one of each. This allowed us to consider that same set of users for all three attributes. 
 
 For egocentric network analysis, we  only included  users who had at least 100 friends so the distribution of FSDs would be measured over  a reasonably large sample. 

Our final dataset had 78,227 users. For 20,988 users, we also had their egocentric networks with friend, follower, and status counts for all the people they were following.

\subsection{Google Plus}
The Google Plus network  \cite{leskovec2012learning} is part of the Stanford Network Analysis Project (SNAP) datasets, The social network is provided as an adjacency list. We made one pass through the ``combined'' dataset, counting the number of friends a person had. The network is directed, so we counted outgoing edges. In addition to using the friend counts for each person, we were able to get the friend count for each of their friends, thus allowing us to construct a FSD distribution for each egocentric social network.

After processing, we had data for 19,540 users.

\subsection{Pinterest}
The Pinterest data was provided as part of the Social Curation Dataset \cite{www2014zhong}. We used the ``Pinterest User Information'' data, which contained follower, following, and pin (i.e. post) counts. After filtering out people with no followers or pins, we had data for 39,586,033 users in our sample. 

\subsection{Live Journal}
Live Journal Dataset \cite{backstrom2006group,DBLP0810-1355} is one of the SNAP datasets.  We followed the same processing procedure as we used for the Google Plus dataset. After processing, we had data for 44,886 users.

\section{Results}
We found that the distribution of FSD among friends in all five datasets closely followed the values expected from Benford's law, with one interesting exception in the Pinterest following relationship. We will set that exception aside for now and discuss it later.

For Facebook, Google Plus, and LiveJournal friends, Twitter friends and followers, and Pinterest followers, all the distributions of FSDs followed Benford's Law. 

Note that with datasets of this size, it is not appropriate to conduct a statistical hypothesis test for goodness of fit, since over tens of thousands or millions of people, even a very tiny deviation would cause us to reject the null hypothesis. Furthermore, conformance with Benford's Law has never been about a perfect statistical match to the predicted values, even in Benford's original work on the subject \cite{benford1938law,diaconis1979rounding}. Rather, the relative frequencies of FSDs are the guiding principle.

Figure 1 shows the distribution of FSDs for each of the six datasets. 
\begin{figure}
\begin{center}
\includegraphics[width=3.5in]{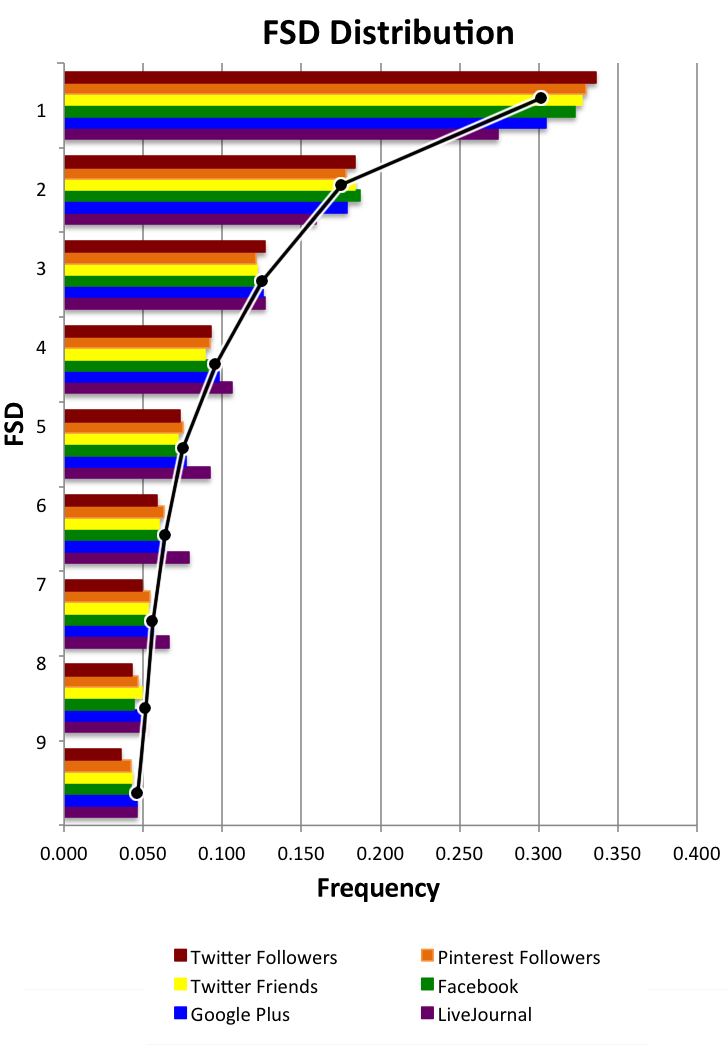}
\caption{Distribution of first significant digits for  Twitter (friends and followers), Google Plus, Pinterest followers, Facebook, and LiveJournal. The black trendline shows the  value predicted by Benford's Law for each FSD.}
\label{fig:dist}
\end{center}
\end{figure}

Pearson correlations are a common way to measure how closely a distribution adheres to Benford's Law \cite{judge2009detecting}. The correlation between the FSDs of the friend/follower counts and what Benford's Law predicts are extremely strong.  As shown in Table 2, all the $r$ values are greater than 0.990.

\begin{table}
\begin{centering}
\label{t:correl}
\caption{Pearson Correlation between the distribution of first significant digits on friend/follower distributions of various social networks with values predicted by Benford's Law.}
\begin{tabular}{lrr} 
\rowcolor{Gray}
\textbf{Site} & \textbf{Total Users} & \textbf{Correlation}\\
\textbf{Google Plus} & 19,500 & 0.9999\\
\rowcolor{LGray}\textbf{Facebook} & 18,298 & 0.9996\\
\textbf{Twitter Friends} & 78,226 & 0.9990\\
\rowcolor{LGray}\textbf{Twitter Followers} & 78,226 & 0.9998\\
\textbf{Pinterest Followers} & 39,586,033 & 0.9989\\
\rowcolor{LGray}\textbf{LiveJournal} & 44,011 & 0.9954

\end{tabular}
\end{centering}
\end{table}

The fit with Benford's law extended to other user behavior. We had data for the number of posts users made on Pinterest (number of pins) and Twitter (number of tweets). In both cases, correlation with Benford's predictions was extremely high: 0.9998 and 0.9960 respectively. 

However, as mentioned above, there was one dataset that did not follow Benford's predictions. On Pinterest, users have both a follower count, which represents incoming social connections, and a following count for outgoing edges. The follower count is what we presented above, and it follows the expected distribution. The \textit{following} count did not adhere to Benford's Law (see figure 2). The percentages are very far off what the law would predict, and the dominance of FSDs of 5 is especially striking. 

Is this simply an exception to the rule, or is something else going on? When Benford's Law is applied in forensic accounting, auditors know to look for explanations of data that appears unusual. For example, a company may have a high percentage of FSDs of 3, not because anything fraudulent is happening, but because they happen to frequently purchase an item that costs \$39.99. 

We investigated this issue on Pinterest more deeply and found the explanation for the frequent 5s. When new users sign up for Pinterest, they are prompted to choose ``interests'' to follow. Users \textit{must} select at least five before continuing with the registration process. This creates at least five initial following relationships for users. Though users can go in and later delete those follows, few do, and this initiation process affects the entire distribution of FSDs. When we looked at the edges in the opposite direction, considering the incoming follower edges instead of the outgoing following edges,  they FSDs adhered to Benford's Law as was shown above. 

This highlights an important point about applying Benford's Law. It can be violated when there is external influence over people's natural behavior. In the Pinterest case, we discovered the influence was an artifact of the system configuration. 

\begin{figure}
\begin{center}
\includegraphics[width=3.5in]{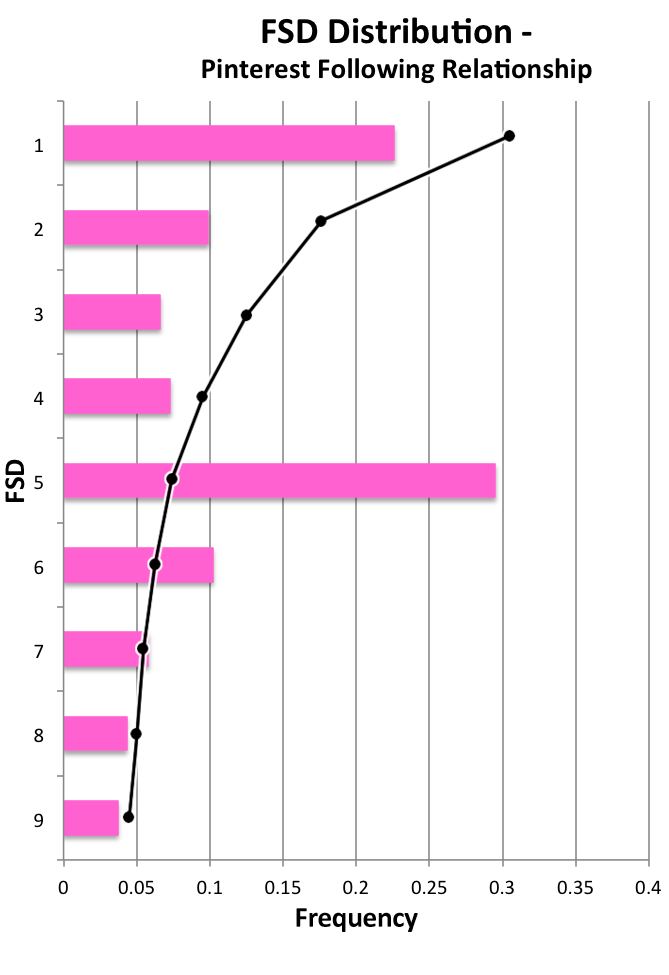}
\caption{Distribution of first significant digits for Pinterest users' following relationships. The black trendline shows the  value predicted by Benford's Law for each FSD.}
\label{fig:pin}
\end{center}
\end{figure}

\subsection{Egocentric Networks}
The adherence to Benford's Law carries through into FSD distributions within individual egocentric networks. Using data from Twitter, Google Plus, and Live Journal, we were selected individuals with at least 100 friends and obtained the number of social connections that each of those friends had. We then determined the distribution of FSDs in the friend-of-friend counts within each egocentric network. Overall, the vast majority of egocentric networks conformed to what Benford's Law predicted.

On Google Plus, 91.5\% of users' egocentric networks' FSD distributions had a correlation of over 0.9 with Benford's Law predictions.  This was true for 85.1\% of LiveJournal egocentric networks. 
 In the Twitter data, 89.7\% of users had a correlation of 0.9 or greater. Only 170  our of 20,988 users ($<1\%$) had a correlation under 0.5. 
 
 Since we had non-anonymized data for Twitter, we were able to look at these accounts with low correlations. Almost every one of the 170 accounts mentioned above appeared to be engaged in suspicious activity. Some accounts were spam, but most were part of a network of Russian bots that posted random snippets of literary works or quotations, often pulled arbitrarily from the middle of a sentence. All the Russian accounts behaved the same way, following other accounts of their type, posting exactly one stock photo image, using a different stock photo image as the profile picture. While we are currently investigating the underlying reason for these bot-controlled accounts to exist, their deviation from Benford's Law made it quite easy to identify their highly unusual behavior.  Only 2 accounts of the 170 seemed like they  belonged to legitimate users.

\section{Discussion and Conclusions}
We have shown that Benford's Law applies to relationships in online social networks. This is true for social networking sites as a whole and for individual users' egocentric networks. Data from Twitter and Pinterest also suggest that it applies to the number of posts users make on social media sites, as well. In the one network where Benford's Law did not hold, further investigation revealed this was due to a feature of the system that altered users' behavior.

There are a number of applications for these results. First, Benford's Law can be used to detect users who are behaving in unexpected ways. As we found in our Twitter dataset, the vast majority of accounts that seriously deviated from the expected distribution of FSDs were engaged in unusual behavior. As is the case with forensic accounting investigations using Benford's Law, a deviation does not necessarily mean there is fraud happening. Given the large number of users on social media, it would be statistically unusual to have no accounts that naturally deviate from expected patterns. Rather, deviation from a Benford distribution can flag accounts for additional review. 

These insights can also be used to validate experimental datasets. It is often the case that data can be hard to collect from social media sites, especially when researchers are looking for detailed personal information. Truly random or representative sampling is difficult to do, and it is essentially impossible when connected components of the social network are important to the analysis. This raises the question as to whether the sample of accounts that a research team collects seriously deviates from normal patterns. While Benford's Law only addresses one aspect of expected behavior, major differences in the FSD distribution on a sample vs. what Benford's Law predicts could indicate serious sampling problems.

We tested this by analyzing the FSD distributions on a number of datasets collected for various projects and experiments. 

We randomly selected 50 Twitter-based networks from the NodeXL Graph Gallery at http://nodexlgraphgallery.org/ . These were all generated by collecting the networks of who had tweeted a given search term. For each graph we analyzed the friend, follower, and tweet counts for the users in each dataset. On all graphs and each of the three measures, the Pearson correlation with Benford's expected values was $>0.990$. This shows that structurally, the networks look like we would expect.

We found similarly strong correlations and agreement with Twitter data collected for a research project that posted a survey on a popular psychology website. Subjects included their Twitter IDs in their survey responses. The distributions of FSDs for friends, followers, and tweets all correlated with Benford's Law distributions with $r>0.990$ and very close values.

However, not all datasets were a good match. One Facebook dataset of 151 users had a Pearson correlation of 0.761, and there were large differences between the predicted and actual frequencies. All but two FSDs saw deviations over 25\% from expected values, and some saw deviations over 80\%. This was true on another Facebook dataset with 220 users supplied by a colleague. In this example, friend counts were self-reported and 94.5\% of those began with a 1 - more than triple the expected 30.1\%.

Such deviations do not necessarily imply there is a problem with the data; indeed, this distribution may be irrelevant to the analysis being performed. However, it hints that the subjects are not reflecting an expected distribution, and thus may vary from the larger population in other ways. Further research is necessary to understand the implications of deviation in experimental samples.

There is a growing understanding of the subtle patterns of natural behavior that humans have difficulty replicating in unnatural circumstances. The applicability of Benford's Law to social media  is a new tool for analyzing user behavior, understanding when and why natural deviations may occur, and ultimately detecting when abnormal forces are at work.

%
\bibliographystyle{plain}
\bibliography{benford}
\end{document}